\documentclass[aps,prd,twocolumn,groupedaddress,showpacs]{revtex4}
\usepackage{graphicx}
\usepackage{dcolumn}
\usepackage{bm}
\usepackage{amssymb,amsmath,latexsym,amsfonts}

\begin{document}

\title{Polymer Bose--Einstein Condensates}

\author{E. Castellanos}
\email{ecastellanos@fis.cinvestav.mx.} \affiliation{Departamento de
F\'isica, Centro de Investigaci\'on y de Estudios Avanzados del
Instituto Polit\'ecnico Nacional, A. P. 14-740, 07000 M\'exico D.
F., M\'exico.}
\author{G. Chac\'{o}n-Acosta}
\email{gchacon@correo.cua.uam.mx} \affiliation{Departamento de
Matem\'aticas Aplicadas y Sistemas, Universidad Aut\'onoma
Metropolitana-Cuajimalpa, Artificios 40, M\'exico D. F. 01120,
M\'exico}

\begin{abstract}
In this work we analyze a non--interacting one dimensional polymer Bose--Einstein condensate in an harmonic trap within the semiclassical approximation. We use an effective Hamiltonian coming from the polymer quantization that arises in loop quantum gravity.
We calculate the number of particles in order to obtain the critical temperature. The Bose--Einstein functions are replaced by series, whose high order terms are related to powers of the polymer length.
It is shown that the condensation temperature presents a shift  respect to the standard case, for small values of the polymer scale. In typical experimental conditions, it is possible to establish a bound for  $\lambda^{2}$ up to $  \lesssim 10 ^{-16}$m$^2$. To improve this bound we should decrease the frequency of the trap and also decrease the number of particles.
\end{abstract}

\date{\today}
\maketitle

\section{Introduction}

Several quantum gravity models suggest that the relation between energy and momentum of microscopic particles must be modified as a consequence of the quantum structure of space--time \cite{ami,Giovanni1,Claus,Claus1,5,Kostelecky,amelino1,12,13}. A deformed dispersion relation, emerges as an ideal tool in the search for possible effects related to the quantum structure of space--time. Nevertheless, the most difficult aspect in the search of experimental evidence relevant for the quantum gravity problem is the smallness of the possible effects \cite{Kostelecky,amelino1}. Unfortunately, due to this fact, the possible bounds
for the deformation parameters, open a wide range of
possible magnitudes, which implies a significant challenge.

Among the models that introduce such deformations, polymer quantum systems are simple quantum mechanical models quantized in a similar way as in loop quantum gravity \cite{ashpqm,CVZ1}. These systems provide scenarios where some characteristics of the full theory, specially the discrete nature of space, can be explored in a simpler context \cite{Winkler,kun,chiou,PUR,TPQM}. Indeed, in polymer quantum mechanics the momentum operator can not be defined, hence a regularized operator is proposed by the introduction of the so-called polymer length scale \cite{ashpqm,CVZ1}. Particularly, from this model an effective Hamiltonian which contains some trace of the granularity of space can be drawn \cite{chiapas}.

The use of Bose--Einstein condensates as a tool in the search of quantum gravity effects, for instance, in the context of Lorentz violations, or to provide phenomenological constrains on Planck scale physics, has produced several interesting research \cite{Claus,Claus1,Colladay,Donald,Camacho,CastellanosCamacho,CastellanosCamacho1,CastellanosClaus,Castellanos,Castellanos1}
 (and references therein). In these works, the  possible effects arising from Planck scale physics by looking at some modifications in the thermodynamic properties of Bose-Einstein condensates was explored. Since its observation with the help of magnetic traps \cite{Anderson,Davis,CC,CC1}, the phenomenon of Bose--Einstein condensation, from the experimental and theoretical point of view, has spurred an enormous amount of publications \cite{bagnato,Giorgini,Dalfovo,Ketterle,Haugerud,Li,zijun,Salasnich,Zobay,
Jaouadi,Politzer,RP,Anderson,grossmann,grossmann1,grossmann2,Andrews,Andrews2,yukalov,yukalov1,yukalov3,Vitaly,pathria,Pethick} (and references therein).
In particular the condensates have been studied in different spatial dimensions \cite{yan,Dalfovo,Karabulut,Pethick}. Although, a one dimensional condensate can never be reached, it is possible to obtain a quasi one dimensional condensate, by using extremely anisotropic traps  \cite{Dalfovo,Ketterle,GO,SC}. Additionally, the condensation process of a bosonic gas in one dimension, trapped in a box or in an harmonic oscillator potential, has a very particular behavior. In this situation, apparently the condensation is not possible in the thermodynamic limit if one assumes that the ground state energy is zero, or equivalently, if the associated chemical potential is zero at the condensation temperature. However, taken into account the minimum energy associated to the system, the condensation process is possible at finite temperature \cite{yukalov1,Pethick,Dalfovo,Ketterle}.

The aim of this work is to analyze the properties of a polymer one dimensional condensate trapped in a harmonic oscillator potential, within the semiclassical approximation, in order to obtain representative bounds for the polymer length scale. We will show also that for this one dimensional system the functional form of the relevant thermodynamic quantities, particularly, for the number of particles, can be expressed by series of the so--called Bose--Einstein functions. This series converges when the polymer scale is small enough. Finally, by the introduction of the ground state energy associated to this system, we calculate the shift in the usual condensation temperature caused by the polymer scale $\lambda^2$. These facts allow us to bound the polymer scale by using one dimensional finite size systems.

\section{Polymer Quantization}

Let us now present some of the main results of polymer
quantization \cite{ashpqm,CVZ1}. This quantization arises from loop quantum gravity \cite{vel}.
In the \textit{loop or polymer}
representation the corresponding Hilbert space $\mathcal{H}_{poly}$ is
spanned by the basis states $\{|x_j\rangle\}$, whose coefficients have a suitable
fall-off \cite{ashpqm}, with the following inner product
\begin{equation}\label{IP}
    \langle x_i|x_j\rangle  = \delta_{i,j},
\end{equation}
where $\delta_{i, j}$ is the Kronecker delta. The polymer Hilbert space can be represented
as $\mathcal{H}_{\textrm{poly}}=L^2(\mathbb{R}_d,\textrm{d}\mu_{d})$,
where $d\mu_d$ is the corresponding Haar measure, being $\mathbb{R}_d$ the
real line endowed with the discrete topology, that is, the dual of the
Bohr compactification of the real line \cite{ashpqm,vel}. The basic operators in this quantization are the position and translation. The position operator $\hat{x}$ acts as usual by multiplication
$$
\hat{x}|x_j\rangle=x_j|x_j\rangle\,,
$$
while the translation operator $\hat{V}(\lambda)$ moves to a position of arbitrary distance $\lambda$
$$
\hat{V}(\lambda)|x_j\rangle=|x_j-\lambda\rangle.
$$
In the Schrodinger quantization, the operator $\hat{V}(\lambda)$ is weakly continuous in $\lambda$ and the momentum operator is defined as its infinitesimal generator. However, in the polymer representation the translation operator $\hat{V}(\lambda)$ fails to achieve this condition, i.e., it is not weakly continuous
on $\lambda$ due to the discrete structure assigned to space \cite{ashpqm}. Therefore, there is no Hermitian operator as infinitesimal generator of translations and thus, the momentum operator is not well defined.
Due to this fact, any phase space function has to be regularized by the introduction of a lattice of fixed positive length. We interpret this length as a fundamental scale that is called the polymer length scale  \cite{ashpqm}.
With this regularization, the polymer Hamiltonian can be defined as follows
\begin{equation}\label{ham}
  \widehat{H}_{\lambda} = \frac{\hbar^2}{2m\lambda^2} \left[ 2- \hat{V}(\lambda) -\hat{V}(-\lambda)\right]+ \hat{U}(x),
\end{equation}
where $\hat{U}(x)$ is the potential term. The action of the Hamiltonian
(\ref{ham}) decomposes the polymer Hilbert space
$\mathcal{H}_{\textrm{poly}}$, into a continuum of separable superselected subspaces, each with support on a regular lattice $ \{n \lambda + x_0|n\in \mathbb{Z}\}$, with $x_0\in[0,\lambda)$ that parameterizes a particular superselected sector \cite{ashpqm}.
Notice that the Hamiltonian (\ref{ham}) can be formally written as
\begin{equation}\label{ham-sin}
    \frac{\hbar^{2}}{2m\lambda^2}\widehat{\sin^{2}\Bigl(\frac{\lambda
p}{\hbar}\Bigr)} + \hat{U}(x).
\end{equation}
We can use this expression to obtain the effective Hamiltonian, simply by replacing the kinetic term with the square of the sine function.

It is worth mentioning, as was pointed out in \cite{chiapas}, that the effective three dimensional case present some complications in calculating analytically the integrals. Due to this fact, in the present work we restrict ourselves to the one--dimensional case.

\section{Condensation Temperature}

Let us start with a one--dimensional effective polymer Hamiltonian \cite{chiapas}, given by
\begin{equation}
H=\frac{\hbar^{2}}{2m\lambda^2}\sin^{2}\Bigl(\frac{\lambda
p_{x}}{\hbar}\Bigr) + U(x) \label{Ham1}
\end{equation}
where $U(x)=m\omega_{x}^{2} x^{2}/2$ is the one dimensional
harmonic oscillator potential that model the trap, and $\lambda$ is the so-called polymer
length scale. Therefore, the semiclassical energy spectrum
associated to (\ref{Ham1}) can be expressed as follows
\begin{equation}\label{ep}
\epsilon_{p}=\frac{\hbar^{2}}{2m\lambda^2}\sin^{2}\Bigl(\frac{\lambda
p_{x}}{\hbar}\Bigr)+\frac{m\omega_{x}^{2}x^{2}}{2}.
\end{equation}
The associated one--dimensional
spatial density can be written as \cite{Dalfovo,Pethick}
\begin{equation}\label{n}
n(x)=\frac{1}{2\pi \hbar}\int
\frac{dp_{x}}{e^{\beta(\epsilon_{p}-\mu)}-1},
\end{equation}
where, as usual, $\beta=1/\kappa T$, being $\kappa$ the Boltzmann
constant, and $\mu$ is the corresponding chemical potential.
Consequently, the total number of particles of the
system is
\begin{equation}
N=\int n(x) dx.
\end{equation}

Let us calculate the spatial density $n(x)$ for the polymer case. To
do so we need to calculate the integral (\ref{n}) substituting the energy
(\ref{ep})
\begin{equation}\label{np1}
n(x)=\frac{1}{2\pi \hbar}\int
\frac{dp_{x}}{Z^{-1}(x)\exp\left(\frac{\beta\hbar^{2}}{2m\lambda^2}\sin^{2}\Bigl(\frac{\lambda
p_{x}}{\hbar}\Bigr)\right)-1},
\end{equation}
where $Z(x) =
\exp\left[{\beta\left(\mu-\frac{m\omega_{x}^{2}x^{2}}{2}\right)}\right]$. The integrand of (\ref{np1}) can be replaced by a
geometric series
\begin{equation}
n(x)=\frac{1}{2\pi \hbar}\int
\sum_{j=1}^{\infty}\left(Z(x)\,e^{-\frac{\beta\hbar^{2}}{2m\lambda^2}\sin^{2}\Bigl(\frac{\lambda
p_{x}}{\hbar}\Bigr)} \right)^j\,dp_{x}.
\end{equation}
The sine function in the argument of the exponential can be written
by using an identity, then we interchange the summation and the
integration
\begin{equation}
n(x)=\frac{1}{2\pi \hbar} \sum_{j=1}^{\infty} Z^j(x) \int
\,e^{-\frac{j\beta\hbar^{2}}{4m\lambda^2}\left[1 -
\cos\Bigl(\frac{2\lambda p_{x}}{\hbar}\Bigr) \right]
 } \,dp_{x}.
\end{equation}
The last integral can be recognized as a modified Bessel function of first
kind \cite{abrwtz}, if we introduce a
regulator for the integral that corresponds to the polymer length
$\lambda$ \cite{TPQM}, with the result
\begin{equation}\label{np2}
n(x)=\frac{1}{2\pi \hbar}\frac{\hbar \pi}{\lambda}
\sum_{j=1}^{\infty} Z^j(x) \,e^{-\frac{j\beta\hbar^{2}}{4m\lambda^2}}
I_0\left( \frac{j\beta\hbar^{2}}{4m\lambda^2} \right).
\end{equation}

We obtain the total number of particles by
integrating (\ref{np2}) over all space. The only
$x$--dependent function is $Z(x)$, and then the corresponding
integration is straightforward
\begin{equation}
N = \frac{1}{2} \sqrt{\frac{2\pi}{m\beta \omega_x^2}}
\sum_{j=1}^{\infty}
\,\frac{e^{-\frac{j\beta\hbar^{2}}{4m\lambda^2}}}{\lambda} I_0\left(
\frac{j\beta\hbar^{2}}{4m\lambda^2} \right)
\,\frac{e^{j\mu\beta}}{j^{1/2}}.
\end{equation}

When $\lambda \ll 1$, we can use the asymptotic expansions of the modified Bessel functions \cite{gradsh,abrwtz}
\begin{equation}\label{series}
    I_0(u) \approx \frac{e^{u}}{\sqrt{2\pi u}} \sum_{k=0}^{\infty}\frac{(-1)^k}{(2u)^k k!}\frac{\Gamma{\left(k+\frac{1}{2}\right)}}{\Gamma{\left(\frac{1}{2}-k\right)}}.
\end{equation}
Consequently, the total number of particles is given approximately by
\begin{equation}\label{n-serie}
N = \frac{\kappa T}{\hbar \omega_x}\sum_{k=0}^{\infty} g_{k+1}(z)\, \left(-\lambda^{2}\right)^{k} \left(\frac{2m\kappa T}{\hbar^2}\right)^k\frac{\Gamma \left(k+\frac{1}{2}\right)}{k!\Gamma \left(\frac{1}{2}-k\right)},
\end{equation}
where $g_{\nu}(z)$ are the so--called Bose--Einstein functions, being $z=e^{\beta \mu}$ the fugacity \cite{pathria}. Series (\ref{n-serie}) is not convergent for arbitrary values of the polymer length. However, the approximation (\ref{series}) is valid only for $\lambda \ll 1$. We notice that the leading term turns to be  $g_1(z)$, the standard result \cite{Dalfovo,Pethick}. The second term can be considered as a correction of order $\lambda^2$, and is proportional to $g_2(z)$
\begin{equation}\label{npoly}
N = \frac{\kappa T}{ \hbar\omega_x}
\left[g_1(z) +  \lambda^2 \frac{m\kappa T}{2\hbar^2}g_2(z) +  \mathcal{O}(\lambda^4) \right].
\end{equation}

It is well know that a one--dimensional
condensate cannot exist in the thermodynamic limit, due to the
divergent behavior of the Bose--Einstein function $g_{1}(1)$.
Nevertheless, if we take into account the ground state energy of the
system, we are able to obtain a well defined condensation temperature
\cite{Dalfovo,Pethick,yukalov}. The ground state energy associated
with our system was already calculated in \cite{ashpqm,TPQM,PUR}
and is given by
\begin{equation}
\epsilon_{0}=\hbar \omega_x \frac{d^2}{\lambda^2}\left(1+\frac{\lambda^4}{8d^4}a_0\left(\frac{4d^4}{\lambda^4}\right)\right),
\end{equation}
where $d^2=\hbar/m\omega_x$, is the characteristic length of the one--dimensional oscillator, and $a_n(x)$ is the Mathieu characteristic function \cite{abrwtz}. However, in the limit $\lambda \ll 1$ one can regain the usual ground state energy plus corrections caused by the polymer scale
\begin{equation}\label{e0}
\epsilon_{0}= \frac{\hbar \omega_x}{2} - \frac{\lambda^2}{32}m \omega_x^2 + \mathcal{O}(\lambda^4).
\end{equation}
At the condensation temperature, the value of chemical potential $\mu$ takes the minimum energy associated to the system. Substituting (\ref{e0}) into (\ref{npoly}) and by using the properties of Bose--Einstein functions when $\epsilon_0 / \kappa T \ll1$, \cite{pathria}
we can reexpress (\ref{npoly}) at the condensation temperature as follows
\begin{equation}\label{nc}
N = \frac{\kappa T_c}{\hbar\omega_x}
\left[g_1(z_c) +  \lambda^2 \frac{m\kappa T_c}{2\hbar^2}g_2(z_c) +  \mathcal{O}(\lambda^4) \right],
\end{equation}
where $z_c=e^{-\epsilon_0/\kappa T_c}$.

Setting $\lambda=0$ we recover the usual expression for the number of particles \cite{Dalfovo,Ketterle}
\begin{equation}\label{n1d}
N = \frac{\kappa T_0}{\hbar\omega_x} \ln\left(\frac{2\kappa
T_0}{\hbar\omega_x}\right),
\end{equation}
where $T_{0}$ is the usual condensation temperature in one dimension without polymer modifications. By using an iterating procedure \cite{yukalov1}, the condensation temperature turns to be
\begin{equation}
\label{T0}
\kappa T_0 \approx \hbar\omega_x \frac{N}{\ln2N}.
\end{equation}
Bose--Einstein condensates in one or two dimensions have been extensively studied \cite{Dalfovo,Pethick,Ketterle} (and references therein). The one dimensional condensation is seemingly not possible in the thermodynamic limit, in other words, the condensation temperature tends to zero when the number of particles tends to infinity \cite{Dalfovo,Pethick,Ketterle,GO,SC}. Nevertheless, if one takes into account the associated ground energy, the condensation is possible at finite temperature. Finite size effects are needed in one dimensional systems, in order to make the condensation possible.

In order to calculate the condensation temperature $T_c$, from (\ref{nc}), we introduce the following well known expressions for the Bose--Einstein functions \cite{pathria}
\begin{eqnarray}
  g_1(-e^{\epsilon_0/\kappa T_c} ) &\approx& \ln (\kappa T_c/ \epsilon_0), \\
  g_2(-e^{\epsilon_0/\kappa T_c} ) &\approx& \zeta(2) - \frac{\kappa T_c}{\epsilon_0}\left[ 1 + \ln\left(\frac{\kappa T_c}{\epsilon_0}\right) \right].
\end{eqnarray}
After some algebraic manipulation, we are able to express the number of particles as follows
\begin{eqnarray}
N &=& \frac{\kappa T_c}{\hbar\omega_x} \ln\left(\frac{2\kappa
T_c}{\hbar\omega_x}\right) \left(1 -
\frac{\lambda^2}{2d^2} \right) \nonumber \\
&& -\frac{7}{8}\pi  \frac{\lambda^2}{\Lambda^2_{c}}\left(1-
\zeta(2)\frac{8}{7}\frac{\kappa T_c}{\hbar\omega_x} \right)+
\mathcal{O}(\lambda^4),\label{num2}
\end{eqnarray}
where $\Lambda^2_{c}= 2\pi\hbar^2/ m\kappa T_c$, is the standard one dimensional thermal wave length.
Notice that the leading term is identical to (\ref{n1d}), with $T_{c}$ instead of $T_{0}$. From (\ref{num2}), we can notice that there are two kinds of corrections to the number of particles, associated with the polymer length. One is due to the ratio $\lambda^2/d^2$, that is related to the effective size of the ground state of the harmonic trap. The second kind of correction is due to $\lambda^2/\Lambda^2$, this term can be interpreted as a pseudo--interaction within the system \cite{CastellanosCamacho1}.

To calculate the shift caused by the polymer scale $\lambda^{2}$,
let us expand (\ref{num2}) at first order in $\lambda^{2}=0$ and $T_{c}=T_{0}$, recalling that $T_{0}$ is the usual condensation temperature given by equation (\ref{T0}). The resulting shift $\Delta T_c$ is
\begin{equation}
\label{shift}
\frac{T_{c}-T_{0}}{T_{0}}\equiv\frac{\Delta T_{c}}{T_{0}}\approx -\frac{\lambda^{2}}{2d^{2}}\Bigg[\frac{\ln 2N-\frac{7}{8}+\frac{\zeta(2)N}{\ln 2N}}{1+ \ln 2N}\Bigg]
\end{equation}

Notice that the possibility to obtain a measurable co\-rrection associated to the polymer scale $\lambda^{2}$ ($\delta T_{c}^{\lambda^{2}}$) requires that, if $\Delta
(T_{c})$ is the experimental error, then $\Delta (T_{c}) \lesssim \vert \delta T_{c}^{\lambda^{2}}\vert$, which in our case this entails
\begin{equation}
\label{Exp}
\Delta (T_{c}) \lesssim \Bigg |\frac{\lambda^{2}}{2d^{2}}\Bigg[\frac{\ln 2N-\frac{7}{8}+\frac{\zeta(2)N}{\ln 2N}}{1+ \ln 2N}\Bigg]\Bigg|.
\end{equation}
In order to obtain representative bounds associated with the polymer scale,  let us appeal to the current high precision experiments
for $^{39}_{19} K$ in a 3--Dim condensate \cite{RP}. Although, from the experimental point of view there is no a real one dimensional condensate, it is possible to construct a quasi one dimensional condensate just by using extremely anisotropic traps \cite{GO}. Then, in principle, the use of this experimental accuracy is justified to bound the polymer scale $\lambda^2$.
The shift in the condensation temperature
respect to the ideal result, caused by the interactions among the
constituents of the gas is about $5\times 10^{-2}$ with a $1\%$  of
error \cite{RP}. For instance, under typical conditions the number of particles is about $N \sim 10^{6}-10^{5}$ in three dimensions \cite{Dalfovo}; for one dimensional systems, the number of particles can be estimated up to  $N \sim 10^{4}$ \cite{GO,SC,Karabulut}. Using typical frequencies of order $100$ Hz, allows us to bound the polymer scale up to $\lambda^{2}  \lesssim 10 ^{-16}$ m$^2$, which is notable. Indeed, as far as we know, this is the first bound associated to $\lambda$, from low energy, earth based experiments.

Notice that, if we increase the number of particles for a fixed trap frequency of order $10^2$Hz, the bound associated to the polymer scale decreases. On the other hand,  for a fixed number of particles of order $\sim 10^4$, again the bound decreases, when increasing $\omega$. These facts suggest, that finite size systems are required  in order to obtain representative bounds for $\lambda$.

\section{Discussion}\label{sec:conclutions}

Quantum gravity phenomenology suggests that the quantum structure of space-time could have some effect on the dynamics of particle motion. These effects can be modeled, for instance, as deformations of dispersion relations. Particularly, the quantization that arises in loop quantum gravity applied to the motion of a quantum particle in one dimension (known as polymer quantization), gives a regularized Hamiltonian together with the introduction of a length parameter known as polymer scale. There are many proposals in the experimental search of these effects, however, is of special interest to study the Bose--Einstein condensates as they experimentally offer high accuracy measurements, which allows to establish bounds to the polymer length scale.

The main goal of this paper was to constrain the polymer length parameter by studying an non--interacting one dimensional polymer Bose--Einstein condensate in an harmonic trap within the semiclassical approximation. We have proved that the corresponding expression (\ref{n-serie}) generalize the so-called Bose--Einstein functions that appears in the standard case, to a series in terms of powers of the polymer length scale. Using  typical experimental values for the number of particles in the condensate, the mass, and the frequency of the harmonic trap, we were able to establish a bound for the polymer length up to $\lambda^2 \lesssim 10^{-16}$m$^2$.
To improve this bound, finite size systems in one dimension are required.

Notice that if we consider the ratio $(\lambda/d)^{2}$ as an effective dimensionless parameter then in principle, we are able to compare with the deformation parameters suggested in other approaches \cite{CastellanosClaus,Castellanos,Castellanos1} by using condensates. We can see that the order of magnitude are quite similar and opens the possibility to relate different scenarios as was recently suggested \cite{MAJ}.

It is possible to consider more general traps like power law potentials, the 3D case, and interactions among the constituents of the gas, to improve further the bound on the polymer scale, and will be presented elsewhere \cite{NOS}.

\begin{acknowledgments}
This work was partially supported by project PROMEP 47510283, CONACyT no. 167563 (GCA).
 E. C. acknowledges CONACyT for the posdoctoral grant received and also
CONACyT M\'exico, CB-2009-01, no. 132400, CB-2011, no. 166212,  and I0101/131/07
C-234/07 of the Instituto Avanzado de Cosmolog\'ia (IAC)
collaboration (http://www.iac.edu.mx/).
\end{acknowledgments}

\end{document}